\documentclass[11pt]{article}
\usepackage{amsfonts,amsmath} 

\newfont{\gl}{eufm10 scaled \magstep1} 

\numberwithin{equation}{section}

\providecommand{\bysame}{\leavevmode\hbox to3em{\hrulefill}\thinspace}
\providecommand{\MR}{\relax\ifhmode\unskip\space\fi MR }

\providecommand{\href}[2]{#2}

\newcommand{\dd}{\hbox{d}}

\begin{document}

\title{\bf{Supersymmetric WZ-Poisson sigma model and twisted
generalized complex geometry}} \author{\ \\Iv\'an
Calvo\footnote{e-mail: {\tt icalvo@unizar.es}}\\ {} \\ Department of
Theoretical Physics\\ University of Zaragoza, E-50009 Zaragoza, Spain}
\date{ } \maketitle

\begin{abstract}
\vskip 0.3cm
\noindent

It has been shown recently that extended supersymmetry in twisted
first-order sigma models is related to twisted generalized complex
geometry in the target. In the general case there are additional
algebraic and differential conditions relating the twisted generalized
complex structure and the geometrical data defining the model. We
study in the Hamiltonian formalism the case of vanishing metric, which
is the supersymmetric version of the WZ-Poisson sigma model. We prove
that the compatibility conditions reduce to an algebraic equation,
which represents a considerable simplification with respect to the
general case. We also show that this algebraic condition has a very
natural geometrical interpretation. In the derivation of these results
the notion of contravariant connections on twisted Poisson manifolds
turns out to be very useful.

\end{abstract}
\vfill\eject

\section{Introduction}

Generalized complex structures (\cite{Hit},\cite{Gua}) on a manifold
$M$ are objects defined on $TM\oplus T^*M$ which unify complex and
symplectic geometry and interpolate between them. Recently, there has
been much activity related to field theory realizations of generalized
complex structures. Recall that extended supersymmetry in
two-dimensional sigma models is related to complex geometry
(\cite{GatHulRoc}) in the target. On the other hand, in the
first-order formulation the fields of the models take values in
$TM\oplus T^*M$. This led to investigate the conditions for the
existence of extended supersymmetry in first-order actions. The task
was started in \cite{Lin} and systematically carried out in
\cite{LinMinTomZab}, where it was shown that extended supersymmetry is
closely related to generalized complex geometry in the target.

Later, in \cite{Zab}, it was shown that the Hamiltonian formalism
allows to rederive and extend the results of \cite{LinMinTomZab} in a
transparent fashion. In particular, it is proved in a
model-independent way that if the generators of a second supersymmetry
(denoted by ${\bf Q}_2(\epsilon)$) are to satisfy the supersymmetry
algebra, the target must be a (twisted, if a WZ term is present)
generalized complex manifold. Choosing a concrete model (i.e. a
concrete Hamiltonian) and imposing that ${\bf Q}_2(\epsilon)$ be a
symmetry should yield some compatibility conditions relating the
geometrical objects defining the model and the generalized complex
structure ${\cal J}$. In the general case (\cite{LinMinTomZab}) these
conditions include a number of algebraic as well as differential
equations.

Following the approach of \cite{Zab} we work out these compatibility
conditions for the case of vanishing metric in the first-order sigma
model, including a WZ term. When the closed 3-form $H$ defining this
twisting term vanishes, it is the supersymmetric version of the
Poisson sigma model (\cite{SchStr94}), related to deformation
quantization (\cite{CatFel99},\cite{CalFal051}). For $H\neq 0$ it is
the supersymmetric version of the WZ-Poisson sigma model
(\cite{KliStr}), whose target is equipped with a twisted Poisson
structure $\Pi$. We show that the notion of contravariant connections
on twisted Poisson manifolds is the key for unraveling the
differential compatibility conditions between $\Pi$ and ${\cal J}$. We
prove that, remarkably, the differential compatibility conditions are
implied by the algebraic ones\footnote{This was already pointed out
for the untwisted Poisson sigma model in \cite{Ber}.}, for which we
also give a simple geometrical interpretation. The main result of this
paper can be stated as follows:
\vskip 0.3cm
The WZ-Poisson sigma model admits $N=2$ supersymmetry if and only if
$${\cal J}(L_\Pi)\subset L_\Pi$$
where $L_\Pi\subset TM\oplus T^*M$ is the Dirac structure associated to
$\Pi$ (see definitions in Section 3).

\section{Contravariant connections on twisted Poisson manifolds}

Let $M$ be an $m$-dimensional manifold. A {\it twisted Poisson
structure} or {\it $H$-Poisson structure} on $M$
(\cite{KliStr},\cite{SevWei}) is defined by a bivector field $\Pi$
and a closed 3-form $H$ which satisfy a modified Jacobi identity:
\begin{equation} \label{Jacobimodified}
\Pi^{\mu\rho}\partial_{\rho}\Pi^{\nu\sigma}+\Pi^{\nu\rho}\partial_{\rho}\Pi^{\sigma\mu}+\Pi^{\sigma\rho}\partial_{\rho}\Pi^{\mu\nu}=\Pi^{\mu\alpha}\Pi^{\nu\beta}\Pi^{\sigma\gamma}H_{\alpha\beta\gamma}
\end{equation}
where $\partial_\mu:=\frac{\partial}{\partial x^\mu}$ and $x^\mu$ are
local coordinates on $M$. Of course, in the untwisted case, $H=0$,
$\Pi$ is a Poisson structure and (\ref{Jacobimodified}) reduces to the
Jacobi identity.

Assume we want to endow the twisted Poisson manifold $(M,\Pi,H)$ with
a covariant connection. It is natural to demand this connection to be
compatible with $\Pi$, so that the covariant derivative of $\Pi$
vanishes. Already in the untwisted case one can prove (\cite{Vai94})
that a compatible covariant connection exists if and only if the rank
of the Poisson tensor is constant. Much of the interest of Poisson and
twisted Poisson structures resides in the fact that the rank of $\Pi$
can be non-constant, differing in an essential way from symplectic and
twisted symplectic manifolds. Therefore, the notion of covariant
derivative is not appropriate for twisted Poisson manifolds. The
relevant concepts in Poisson manifolds are those of contravariant
derivatives introduced by Vaisman (\cite{Vai94}) and contravariant
connections developed by Fernandes (\cite{Fer}). Next, we extend their
definitions and some results to the case of twisted Poisson manifolds.

A {\it contravariant derivative} on a vector bundle $E$ over the
twisted Poisson manifold $(M,\Pi,H)$ is an operator $\nabla$ such that
for each $\alpha \in \Omega^1(M)$, $\nabla_\alpha$ maps sections of
$E$ to sections of $E$ satisfying:

\vskip 0.3cm

(i) $\nabla_{\alpha_1 + \alpha_2}\psi
=\nabla_{\alpha_1}\psi+\nabla_{\alpha_2}\psi$

(ii) $\nabla_{\alpha}(\psi_1 + \psi_2)=\nabla_{\alpha}\psi_1 +
\nabla_{\alpha}\psi_2$

(iii) $\nabla_{f\alpha}\psi=f\nabla_\alpha\psi$

(iv) $\nabla_\alpha(f\psi)=f\nabla_\alpha\psi+
\Pi^\sharp\alpha(f)\psi$
\vskip 0.2cm
\noindent where $\Pi^\sharp_p:T_p^*M\rightarrow T_pM$ is defined by contraction
with the first index of $\Pi$.

\vskip 0.3cm

A contravariant derivative defines a contravariant connection in an
analogous way to the covariant case. We shall be interested in
defining the contravariant derivative of tensor fields on $M$. To this
end it is enough to have a contravariant derivative on
$E=T^*M$. Take local coordinates $x^\mu$ on $M$ and define the
Christoffel symbols $\Gamma^{\mu\nu}_\rho$ by
\begin{equation}
\nabla_{\dd x^\mu}\dd x^\nu = \Gamma^{\mu\nu}_\rho \dd x^\rho
\end{equation}

\newpage

The contravariant derivative of a tensor field $K$ of type $(p,q)$ is given by
\begin{eqnarray}
\nabla^{\sigma}K^{\mu_1\dots\mu_p}_{\nu_1\dots\nu_q}&=&\Pi^{\sigma\rho}\partial_\rho
K^{\mu_1\dots\mu_p}_{\nu_1\dots\nu_q}-\sum_{a=1}^{p}\Gamma^{\sigma\mu_a}_\rho
K^{\mu_1\dots\rho\dots\mu_p}_{\nu_1\dots\nu_q}+\cr
&+&\sum_{b=1}^{q}\Gamma^{\sigma\rho}_{\nu_b}K^{\mu_1\dots\mu_p}_{\nu_1\dots\rho\dots\nu_q}
\end{eqnarray}

A tensor field $K$ is said {\it parallel} if $\nabla K = 0$. The
relevant result for us is given by the following

{\bf Theorem:} {\it Let $(M,\Pi,H)$ be a twisted Poisson
manifold. Then, there exists a contravariant connection such that
$\Pi$ is parallel.}

\vskip 0.3 cm

{\it Proof:} Let $\{U_i\}$ be an open cover of $M$. Take local
coordinates $x^\mu$ on $U_i$ and define
\begin{equation} \label{localsymbols}
\Gamma^{\mu\nu}_{(i)\rho}=\partial_\rho\Pi^{\mu\nu}-\frac{1}{2}\Pi^{\mu\alpha}\Pi^{\nu\beta}H_{\rho\alpha\beta}
\end{equation}
On $U_i$, $\Pi$ is parallel for the contravariant connection
$\nabla_{(i)}$ with symbols $\Gamma_{(i)}$. If $\sum_i f_i=1$ is a
partition of unity subordinated to the cover $\{U_i\}$, $\nabla=\sum_i
f_i\nabla_{(i)}$ gives a contravariant connection on $M$ such that
$\nabla\Pi=0$.  $\,$\hfill$\Box$\break

\section{The Courant bracket on $TM\oplus T^*M$. Twisted generalized complex structures and twisted Poisson structures.}

Let $M$ be an $m$-dimensional differentiable manifold. Given a closed
3-form $H$ the $H$-{\it twisted Courant bracket} on $TM\oplus T^*M$ is
defined by
\begin{equation} \label{Courant}
[{X}_1+\xi_1,{X}_2+\xi_2]_H=[{X}_1,{X}_2]+{\cal L}_{{X}_1}\xi_2-{\cal
L}_{{X}_2}\xi_1-{\frac{1}{2}}\dd
(i_{{X}_1}\xi_2-i_{{X}_2}\xi_1)+i_{X_1}i_{X_2}H
\end{equation} for
${X}_1+\xi_1,{X}_2+\xi_2\in C^\infty(TM\oplus T^*M)$. It is
skew-symmetric but not a Lie bracket since it does not satisfy the
Jacobi identity. The bracket (\ref{Courant}) makes $TM\oplus T^*M$
into a {\it Courant algebroid}.

\vskip 0.4 cm

An $H$-{\it Dirac structure} $L$ in $TM\oplus T^*M$ is a maximally
isotropic subbundle of $TM\oplus T^*M$ with respect to the natural
pairing
\begin{equation} \label{pairing}
\langle {X}_1+\xi_1,{X}_2+\xi_2 \rangle = \xi_1({X}_2)+\xi_2({X}_1)
\end{equation}
and whose sections are closed under the $H$-twisted Courant bracket.

$H$-Poisson structures can be defined in a nice way in this
language. Namely, the graph of a bivector field $\Pi$,
\begin{equation} \label{Diracstructure}
L_\Pi = \{(\Pi^\sharp\xi,\xi)\in TM\oplus T^*M \vert \xi\in T^*M\}
\end{equation}
is an $H$-Dirac structure if and only if $\Pi$ is an $H$-Poisson
structure.

\vskip 0.4 cm

If $x^\mu$ are local coordinates on $M$ and take $(\partial_\mu,\dd
x^\mu)$ as a basis in the fibres of $TM\oplus T^*M$, the bilinear form
(\ref{pairing}) reads
\begin{equation} \label{metric}
{\cal I}=\begin{pmatrix}0&1_m\cr 1_m&0\end{pmatrix}
\end{equation}

An {\it almost generalized complex structure} is a linear map ${\cal
J}:TM\oplus{T^*M}\rightarrow TM\oplus{T^*M}$ such that ${\cal
J}^t{\cal I}{\cal J}={\cal I}$ and ${\cal J}^2 = -1_{2m}$. For such
${\cal J}$ define $p_{\pm}=\frac{1}{2}(1_{2m}\pm i{\cal J})$. The
almost generalized complex structure ${\cal J}$ is an $H$-{\it twisted
generalized complex structure} if
\begin{equation} \label{integrability}
p_\mp[p_\pm({X}_1+\xi_1),p_\pm({X}_2+\xi_2)]_H=0,\ \forall
{X}_1+\xi_1,{X}_2+\xi_2\in C^\infty(TM\oplus{T^*M})
\end{equation}

\vskip 0.4 cm

In the coordinate basis $(\partial_\mu,\dd x^\mu)$ we can write,
\begin{equation}
{\cal J}=\begin{pmatrix}J&P\cr L&K \end{pmatrix}
\end{equation}
where $J:TM\rightarrow TM,P:T^*M\rightarrow TM,L:TM\rightarrow
T^*M,K:T^*M\rightarrow T^*M$. The condition ${\cal J}^t{\cal I}{\cal
J}={\cal I}$ becomes
\begin{equation}
J^\mu_{\,\,\,\nu} + K_{\mu}^{\,\,\,\nu} =0,\,\,\,\,\,\,\,\,\,\,\,
 P^{\mu\nu} = - P^{\nu\mu},\,\,\,\,\,\,\,\,\,\,\,
 L_{\mu\nu}=-L_{\nu\mu}
\end{equation}
whereas ${\cal J}^2 = -1_{2m}$ translates into
\begin{subequations}
\begin{align}
&J^\mu_{\,\,\,\nu} J^\nu_{\,\,\,\lambda} + P^{\mu\nu} L_{\nu\lambda} =
-\delta^\mu_{\,\,\,\lambda}\label{AA1}\\ &J^\mu_{\,\,\,\nu}
P^{\nu\lambda} + P^{\mu\nu}K_\nu^{\,\,\,\lambda}
=0\label{AA2}\\ &K_\mu^{\,\,\,\nu} K_\nu^{\,\,\,\lambda} +
L_{\mu\nu} P^{\nu\lambda} = -\delta^\mu_{\,\,\,\lambda}\label{AA3}\\
&K_{\mu}^{\,\,\,\nu} L_{\nu\lambda} + L_{\mu\nu} J^\nu_{\,\,\,\lambda}
=0\label{AA4}
\end{align}
\end{subequations}

We shall not need the coordinate expression of the differential
conditions coming from (\ref{integrability}). We just want to point
out that one of these conditions implies that $P$ is a Poisson
structure, i.e.
\begin{equation} \label{JacobiP}
P^{\mu\rho}\partial_{\rho}P^{\nu\sigma}+P^{\nu\rho}\partial_{\rho}P^{\sigma\mu}+P^{\sigma\rho}\partial_{\rho}P^{\mu\nu}=0
\end{equation}

\vskip 0.4 cm

The action of a 2-form $b$ on $TM\oplus T^*M$ defined by
\begin{equation} \label{Btransform}
e^b({X}+{\xi}):= {X}+\xi + i_{X}b
\end{equation}
is said a {\it $b$-transform}. The interesting point is that this
action is a morphism of Courant algebroids. Concretely,
\begin{equation}
e^b([{X}_1+\xi_1,{X}_2+\xi_2]_H)=[e^b({X}_1+\xi_1),e^b({X}_2+\xi_2)]_{H+\dd
b}
\end{equation}

It follows that the $b$-transform of an $H$-twisted generalized complex
structure ${\cal J}$,
\begin{equation}
{\cal J}_b:=e^b{\cal J}e^{-b}
\end{equation}
is an $(H+\dd b)$-twisted generalized complex structure.

Under a $b$-transform (\ref{Btransform}), the $H$-Dirac structure
$L_\Pi$ is transformed into $e^b(L_\Pi)$, which is an $(H+\dd
b)$-Dirac structure. However, the subbundle $e^b(L_\Pi)$ is the graph
of a bivector field if and only if $1_m+b\Pi$ is invertible. In this
case, $e^b(L_\Pi)=L_{\Pi_b}$, where
\begin{equation}
{\Pi_b}:=\Pi(1_m+b\Pi)^{-1}
\end{equation}
is an $(H+\dd b)$-Poisson structure.

If $b$ is closed the transformation (\ref{Btransform}) gives an
orthogonal automorphism of the twisted Courant bracket. The most
general orthogonal automorphism of the twisted Courant bracket is a
semidirect product of a diffeomorphism of $M$ and a $b$-transform with
$b\in\Omega_{{\rm closed}}(M)$.

\section{Extended supersymmetry in the WZ-Poisson sigma model}

Let $S^{1,1}$ be the supercircle with coordinates
${\bar\sigma}=(\sigma,\theta)$. In local coordinates the cotangent bundle of the
superloop space, $T^*{\bf L}M$, is given by scalar superfields
$\Phi^\mu(\sigma,\theta)$ and spinorial superfields
$S_\mu(\sigma,\theta)$. In components,
\begin{equation}
\Phi^\mu(\sigma) = X^\mu(\sigma)+\theta\lambda^\mu(\sigma),\quad
S_\mu(\sigma)=\rho_\mu(\sigma)+\theta p_\mu(\sigma)
\end{equation}
where $X^\mu$ and $p_\mu$ are bosonic fields.

Assume that $M$ is equipped with a closed 3-form $H$. Then, the
 following 2-form on $T^*{\bf L}M$ is symplectic:
\begin{equation} \label{symplecticform}
\omega=\int_{S^{1,1}}\dd\sigma\dd\theta\left(
\delta\Phi^\mu\wedge\delta S_\mu
-\frac{1}{2}H_{\mu\nu\rho}D\Phi^\mu\delta\Phi^\nu\wedge\delta\Phi^\rho
\right)
\end{equation}
where $\delta$ stands for the de Rham differential and
$D=\partial_\theta - \theta\partial_\sigma$. Since $\omega$ is closed
and non-degenerate, it defines a Poisson bracket on
functions of the superfields which we shall denote by $\{\cdot ,\cdot \}$.

The basic Poisson brackets read:
\begin{eqnarray}
&&\{\Phi^\mu({\bar\sigma}),\Phi^\nu({\bar\sigma}')\}=0\cr
&&\{\Phi^\mu({\bar\sigma}),S_\nu({\bar\sigma}')\}=\delta^\mu_\nu
\delta({\bar\sigma}-{\bar\sigma}')\cr
&&\{S_\mu({\bar\sigma}),S_\nu({\bar\sigma}')\}=H_{\mu\nu\rho}D\Phi^\rho\delta({\bar\sigma}-{\bar\sigma}')
\end{eqnarray}
with $\delta({\bar\sigma}-{\bar\sigma}')$ the superspace delta
distribution. Notice that
\begin{equation} \label{canonicaltransformation}
S_\mu\mapsto S_\mu - B_{\mu\nu}D\Phi^\nu
\end{equation}
is a canonical transformation for closed $B$.

The Hamiltonian formulation of the $N=1$ supersymmetric WZ-Poisson
sigma model (WZ-PSM) is as follows. The phase space of the theory,
denoted by ${\cal C}$, is the set of points of $T^*{\bf L}M$
satisfying the constraints:
\begin{equation} \label{constraint}
D\Phi^\mu(\sigma,\theta) + \Pi^{\mu\nu}(\Phi^\mu)S_\nu(\sigma,\theta)
= 0,\ \mu = 1,\dots,m
\end{equation}
which are a consequence of the singular nature of the Lagrangian of
the WZ-PSM (\cite{KliStr}). The Hamiltonian of the WZ-PSM can be
written:
\begin{equation} \label{Hamiltonian}
{\cal H} =
\int_{S^{1,1}}F_\mu(\sigma,\theta)\left(D\Phi^\mu(\sigma,\theta) +
\Pi^{\mu\nu}(\Phi^\mu)S_\nu(\sigma,\theta)\right) \dd\sigma\dd\theta
\end{equation}
where the fields $F_\mu$ act as Lagrange multipliers. The consistency
of the model requires $\Pi$ to be an $H$-Poisson structure. In the
Hamiltonian formalism this is obtained as the condition for the
dynamics to preserve the submanifold ${\cal C}$, i.e. $\{D\Phi^\mu +
\Pi^{\mu\nu}S_\nu,{\cal H}\}\vert_{\cal C}=0$.

By construction, the WZ-PSM is invariant under the supersymmetry
transformation generated by:
\begin{equation}
{\bf Q}_1(\epsilon)=\int_{S^1}\dd\sigma\epsilon \left(
p_\mu\lambda^\mu-\rho_\mu\partial X^\mu
-\frac{1}{3}H_{\mu\nu\rho}\lambda^\mu\lambda^\nu\lambda^\rho \right)
\end{equation}
where $\partial X^\mu = \frac{\partial X^\mu}{\partial\sigma}$.

We address now the issue of extended supersymmetry in the WZ-PSM. The
most general ansatz for a second supersymmetry transformation is given
by (\cite{Zab})
\begin{equation} \label{Q2}
{\bf Q}_2(\epsilon) = -\frac{1}{2}\int_{S^{1,1}}\dd\sigma\dd\theta \epsilon
(2D\Phi^\mu S_\nu J^\nu_{\,\,\,\mu}+D\Phi^\mu D\Phi^\nu
L_{\mu\nu}+S_\mu S_\nu P^{\mu\nu})
\end{equation}

As shown in \cite{Zab} in a model-independent way, the generators
${\bf Q}_2(\epsilon)$ satisfy the supersymmetry algebra if and only if
\begin{equation}
{\cal J}=\begin{pmatrix}J&P\cr L&-J^t \end{pmatrix}
\end{equation}
is an $H$-twisted generalized complex structure. In this context, the
canonical transformation (\ref{canonicaltransformation}) is identified
with a $b$-transform for closed $B$, i.e. an automorphism of the
Courant algebroid. If $B$ is not closed, $S_\mu\mapsto S_\mu -
B_{\mu\nu}D\Phi^\nu$ is not a canonical transformation and it changes
the twisting term in (\ref{symplecticform}). In particular, if $H=\dd
B$ it untwists the symplectic structure.

It remains to find out when ${\bf Q}_2(\epsilon)$ generates a symmetry
transformation of the WZ-PSM. That is, when
\begin{equation} \label{symmetry}
\{{\bf Q}_2(\epsilon),{\cal H}\}\vert_{\cal C}=0
\end{equation}

At this stage one expects that (\ref{symmetry}) holds only if some
compatibility conditions relating ${\cal J}$ and $\Pi$ are
satisfied. These conditions were worked out in the Lagrangian
formalism for a general (untwisted) first-order sigma model in
\cite{LinMinTomZab}. It turns out that in the general case some
algebraic as well as differential conditions must be imposed. Our aim
is to prove that in the WZ-PSM the differential conditions are
automatically implied by the algebraic ones. We shall see that the
contravariant connections introduced in Section 2 are extremely
helpful in the derivation of this result.

A direct calculation shows that (\ref{symmetry}) holds if and only if
the following two conditions are met:

{\it Algebraic condition:}
\begin{equation} \label{algebraic}
P^{\mu\nu}+J^\mu_{\,\,\,\rho}\Pi^{\rho\nu}+\Pi^{\mu\rho}J^\nu_{\,\,\,\rho}-\Pi^{\mu\rho}L_{\rho\sigma}\Pi^{\sigma\nu}=0
\end{equation}

{\it Differential condition:}
\begin{eqnarray} \label{differential}
&&\frac{1}{2}\Big[\Pi^{\mu\beta}\partial_\beta
  P^{\nu\rho}-\partial_\beta
  \Pi^{\mu\nu}P^{\beta\rho}-\partial_\beta\Pi^{\mu\rho}P^{\nu\beta}+\cr
  &&+(\Pi^{\mu\beta}\partial_\beta
  J^\nu_{\,\,\,\gamma}-\partial_\beta\Pi^{\mu\nu}J^\beta_{\,\,\,\gamma}+\partial_\gamma\Pi^{\mu\beta}J^\nu_{\,\,\,\beta})\Pi^{\gamma\rho}+\cr
  &&+\Pi^{\nu\gamma}(\Pi^{\mu\beta}\partial_\beta
  J^\rho_{\,\,\,\gamma}-\partial_\beta\Pi^{\mu\rho}J^\beta_{\,\,\,\gamma}+\partial_\gamma\Pi^{\mu\beta}J^\rho_{\,\,\,\beta})-\cr
  &&-\Pi^{\nu\gamma}(\Pi^{\mu\beta}\partial_\beta
  L_{\gamma\alpha}+\partial_\gamma\Pi^{\mu\beta}L_{\beta\alpha}+\partial_\alpha\Pi^{\mu\beta}L_{\gamma\beta})\Pi^{\alpha\rho}\Big]+\cr
  &&+(P^{\alpha\nu}+J^\alpha_{\,\,\,\tau}\Pi^{\tau\nu})\Pi^{\mu\gamma}\Pi^{\beta\rho}H_{\beta\alpha\gamma}=0
\end{eqnarray}

The untwisted version of the differential condition
(\ref{differential}) was deduced in \cite{LinMinTomZab} (see equation
(6.35) therein) for a general untwisted sigma model. In the case of
non-vanishing metric it was interpreted as a condition of constancy
with respect to a covariant derivative compatible with the metric
tensor. The WZ-PSM is the limit of vanishing metric of the twisted
first-order sigma model and there is no such covariant derivative at
hand. We have learnt in Section 2 that the natural objects which allow
to compare tangent spaces at different points of a twisted Poisson
manifold are contravariant connections. In fact, after some
manipulations\newpage
we can rewrite condition (\ref{differential}) in
terms of the local symbols (\ref{localsymbols})\footnote{We omit the
subscript $(i)$ referring to an open set of a cover of $M$.}:
\begin{eqnarray}
&&\nabla^\mu P^{\nu\rho}+(\nabla^\mu
J^\nu_{\,\,\,\gamma})\Pi^{\gamma\rho}+\Pi^{\nu\gamma}(\nabla^\mu
J^\rho_{\,\,\,\gamma})-\Pi^{\nu\gamma}(\nabla^\mu
L_{\gamma\alpha})\Pi^{\alpha\rho}+\cr
&&+(P^{\alpha\nu}+J^\alpha_{\,\,\,\tau}\Pi^{\tau\nu}-J^\nu_{\,\,\,\tau}\Pi^{\tau\alpha}-\Pi^{\alpha\delta}L_{\delta\tau}\Pi^{\tau\nu})\Pi^{\mu\gamma}\Pi^{\beta\rho}H_{\beta\alpha\gamma}=0
\end{eqnarray}

Using that $\nabla\Pi=0$ we can go one step further and write:
\begin{eqnarray}
&&\nabla^\mu \left( P^{\nu\rho}+
J^\nu_{\,\,\,\gamma}\Pi^{\gamma\rho}+\Pi^{\nu\gamma}J^\rho_{\,\,\,\gamma}
-\Pi^{\nu\gamma}L_{\gamma\alpha}\Pi^{\alpha\rho} \right)+\cr
&&+(P^{\alpha\nu}+J^\alpha_{\,\,\,\tau}\Pi^{\tau\nu}-J^\nu_{\,\,\,\tau}\Pi^{\tau\alpha}-\Pi^{\alpha\delta}L_{\delta\tau}\Pi^{\tau\nu})\Pi^{\mu\gamma}\Pi^{\beta\rho}H_{\beta\alpha\gamma}=0
\end{eqnarray}
which is obviously satisfied if the algebraic condition
(\ref{algebraic}) holds. Hence, we obtain the remarkable result that
the WZ-PSM has a second supersymmetry (\ref{Q2}) if and only if ${\cal
J}$ is a generalized complex structure and the algebraic condition
(\ref{algebraic}) is satisfied. This gives an enormous simplification
with respect to the general sigma model, in which the differential
condition analogous to (\ref{differential}) is not necessarily implied
by the algebraic ones.

Now, we would like to give a geometrical interpretation of
(\ref{algebraic})\footnote{The condition (\ref{algebraic}) was already
deduced in \cite{Ber}, but a geometrical interpretation was lacking.},
which imposes a compatibility condition between the $H$-Poisson
structure $\Pi$ and the Poisson structure $P$. Define two
endomorphisms of $TM\oplus T^*M$ by
\begin{equation}
\tau_1:=\begin{pmatrix}0 & \Pi \cr 0 & 1_m \end{pmatrix},\quad
\tau_2:= \begin{pmatrix}1_m & -\Pi \cr 0 & 0 \end{pmatrix}
\end{equation}

Notice that $\tau_i^2=\tau_i,i=1,2$. $\tau_1$ projects onto the Dirac
structure $L_\Pi$ (see definition (\ref{Diracstructure})) along $TM$. $\tau_2$
projects onto $TM$ along $L_\Pi$. In particular, ${\rm
Im}(\tau_1)={\rm Ker}(\tau_2)=L_\Pi$, which will be the important
point for us.

In matrix notation the condition (\ref{algebraic}) can be expressed as
\begin{equation} \label{interpretation}
\tau_2 \begin{pmatrix}J & P \cr L & -J^t \end{pmatrix} \tau_1 = 0
\end{equation}
which says that ${\cal J}$ can be restricted to an endomorphism of
$L_\Pi$. That is, the WZ-PSM supports extended supersymmetry if and
only if
\begin{equation}
{\cal J}(L_\Pi)\subset L_\Pi
\end{equation}

Note that the canonical transformation (\ref{canonicaltransformation})
can be viewed as a $b$-transform with $b=-B\in\Omega_{{\rm closed}}(M)$
acting on ${\cal J}$ and $\Pi$. Since a canonical transformation does
not modify the Poisson brackets, one would expect that
(\ref{algebraic}) hold for the transformed objects
\begin{equation}
{\cal J}_{-B} = \begin{pmatrix}J_{-B} & P_{-B} \cr L_{-B} & -J_{-B}^t
\end{pmatrix},\quad \Pi_{-B}=\Pi(1_m-B\Pi)^{-1}
\end{equation}
This is not evident from (\ref{algebraic}). However, the result
follows easily by using that the expression (\ref{interpretation}) is
manifestly invariant under a $b$-transform. But the argument is purely
algebraic so that the result holds even if $B$ is not closed.

The fact that in the WZ-PSM there is only one algebraic compatibility
condition between ${\cal J}$ and $\Pi$, given by (\ref{algebraic}),
should make easier the search of backgrounds admitting extended
supersymmetry. Notice that when trying to solve (\ref{algebraic}) one
can use that it is invariant under a $b$-transform. Therefore, if
$H=\dd B$ and $1_m-B\Pi $ is invertible one can look for solutions of
(\ref{algebraic}) in terms of the untwisted objects and twist back at
the end of the day.

\section{Conclusions}

We have continued the work \cite{Zab}, where it was shown that
extended supersymmetry in phase space for twisted two-dimensional
sigma models requires a twisted generalized complex structure ${\cal
J}$ on the target. We have considered the particular case of the
supersymmetric WZ-Poisson sigma model (WZ-PSM) and analysed the
compatibility conditions relating ${\cal J}$ and the twisted Poisson
structure $\Pi$. The notion of contravariant connections on twisted
Poisson manifolds helps to show that the WZ-PSM admits $N=2$
supersymmetry if and only if
$${\cal J}(L_\Pi)\subset L_\Pi$$
where $L_\Pi\subset TM\oplus T^*M$ is the Dirac structure associated to
$\Pi$. The simplification coming from the absence of differential
conditions makes the WZ-PSM into an interesting mathematical
laboratory and should allow to construct examples of supersymmetric
backgrounds not previously discussed in the literature.

In this paper we have not considered the model defined on a surface
with boundary. First, it would be worth to study in detail the
admissible branes for the non-supersymmetric WZ-PSM, which should
yield a twisted extension of the results of \cite{CalFal04}. When
$H=\dd B$ and $1_{m}-B\Pi$ is invertible, one can untwist the theory
and reduce it to the standard Poisson sigma model. However, for
non-exact $H$ or non-invertible $1_{m}-B\Pi$ there are subtle aspects
which deserve to be investigated. Then, one could turn to the
supersymmetric WZ-PSM to identify the D-branes which preserve
supersymmetry.

Of course, it would be interesting to perform the calculations
presented in this paper for a completely general twisted first-order
sigma model, i.e. with non-vanishing metric. This will be the subject
of further research.

\vskip 0.4cm

\noindent{\bf Acknowledgments:} I thank Fernando Falceto for many
insightful remarks and for a careful reading of the manuscript. I am
also grateful to Izu Vaisman for important comments on the proof of
the theorem of Section 2. This work was supported by MEC (Spain),
grant FPU and grant FPA2003-02948.

\end{document}